\documentclass[9pt,twocolumn,twoside]{pnas-new}

\templatetype{pnasresearcharticle} 

\def\LBCO{LBCO~$1/8$}

\def\mathbi#1{\ensuremath{\textbf{\em #1}}}
\def\QCDW{\ensuremath{\mathbi{Q}_{\text{CDW}}}}
\def\QSDW{\ensuremath{\mathbi{Q}_{\text{SDW}}}}

\title{High-temperature charge density wave correlations in La$_{1.875}$Ba$_{0.125}$CuO$_{4}$ without spin-charge locking}

\author[a,1]{H. Miao}
\author[b]{J. Lorenzana} 
\author[c]{G. Seibold}
\author[d]{Y.Y. Peng}
\author[e]{A. Amorese}
\author[e]{F. Yakhou-Harris}
\author[e]{K. Kummer}
\author[e]{N. B. Brookes}
\author[a]{R. M. Konik}
\author[a]{V. Thampy}
\author[a]{G. D. Gu}
\author[d]{G. Ghiringhelli}
\author[d]{L. Braicovich}
\author[a,1]{M. P. M. Dean}

\affil[a]{Condensed Matter Physics and Materials Science Department, Brookhaven National Laboratory, Upton, New York 11973, USA}
\affil[b]{ISC-CNR, Dipartimento di Fisica, Universit\`a di Roma ``La Sapienza'', P. Aldo Moro 2, 00185 Roma, Italy}
\affil[c]{Institut f\"{u}r Physik, BTU Cottbus, P.O.\ Box 101344, 03013 Cottbus, Germany}
\affil[d]{CNR/SPIN, CNISM and Dipartimento di Fisica, Politecnico di Milano, Piazza Leonardo da Vinci 32, 20133 Milano, Italy}
\affil[e]{European Synchrotron Radiation Facility (ESRF), BP 220, F-38043 Grenoble Cedex, France}

\leadauthor{Miao and Dean} 

\significancestatement{Charge correlations have now been identified in the low-temperature phase of essentially all families of underdoped cuprates, but in two apparently distinct types. The first hosts locked charge and spin stripes of which the material La$_{1.875}$Ba$_{0.125}$CuO$_{4}$ (LBCO~$1/8$) is a paradigmatic example. In the second type, a charge density wave (CDW) exists without any obviously related spin correlations. Here we report the discovery of high-temperature CDW correlations that exist at temperatures above the CDW transition in the canonical striped cuprate \LBCO{}. We find that the high-temperature CDW is decoupled from the fluctuating spin density wave, lessening apparent differences in behavior among different materials and setting important constraints for how we understand their electronic states.}

\authorcontributions{Experiment: H.M., Y.Y.P., A.A., F.Y.-H., K.K., N.B.B., L.B., G.G.\ and M.P.M.D. Theoretical calculations: G.S.\ and J.L. Data analysis and interpretation: H.M., J.L., G.S., R.M.K.\, G.G., L.B. and M.P.M.D. Sample growth and preparation: H.M., V.T.\ and G.D.G. Project planning: H.M., G.G., L.B.\ and M.P.M.D.\ Paper writing: H.M.\ J.L., G.S., G.G., L.B.\ and M.P.M.D.}
\authordeclaration{The authors declare no competing financial interests.}
\correspondingauthor{\textsuperscript{1}To whom correspondence should be addressed. E-mail: hmiao@bnl.gov; mdean@bnl.gov}


\begin{abstract}
Although all superconducting cuprates display charge-ordering tendencies, their low-temperature properties are distinct, impeding efforts to understand the phenomena within a single conceptual framework. While some systems exhibit stripes of charge and spin, with a locked periodicity, others host charge density waves (CDWs) without any obviously related spin order. Here we use resonant inelastic x-ray scattering (RIXS) to follow the evolution of charge correlations in the canonical stripe ordered cuprate La$_{1.875}$Ba$_{0.125}$CuO$_{4}$ (LBCO~$1/8$) across its ordering transition. We find that high-temperature charge correlations are unlocked from the wavevector of the spin correlations, signaling analogies to CDW phases in various other cuprates. This indicates that stripe order at low temperatures is stabilized by the coupling of otherwise independent charge and spin density waves, with important implications for the relation between charge and spin correlations in the cuprates.
\end{abstract}

\doi{\url{www.pnas.org/cgi/doi/10.1073/pnas.1708549114}}

\begin{document}

\verticaladjustment{-2pt}

\maketitle
\thispagestyle{firststyle}
\ifthenelse{\boolean{shortarticle}}{\ifthenelse{\boolean{singlecolumn}}{\abscontentformatted}{\abscontent}}{}

%
\begin{figure*}[tb]
\centering
\includegraphics[width=16 cm]{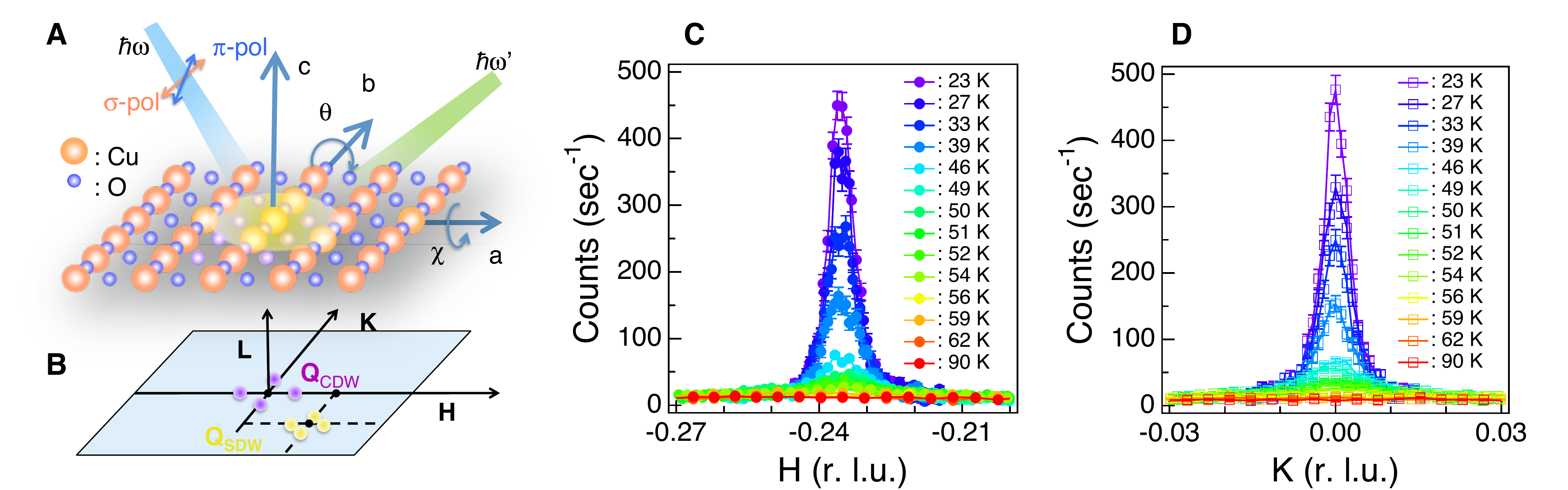}
\caption{\textbf{Scattering geometry and temperature-dependent CDW Bragg peak.} \textbf{A}, The experimental geometry showing incident and outgoing photon directions, labeled by their energies of $\hbar \omega$ and $\hbar \omega^\prime$, scattering from the $c$-axis face of the crystal. The incident x-ray polarization can be tuned to be parallel ($\pi$) or perpendicular ($\sigma$) to the scattering plane. \textbf{B}, The two dimensional (2D) cuprate Brillouin zone. Purple (yellow) points in the 2D Brillouin zone correspond to the locations of charge (spin) density wave Bragg peaks \QCDW{} and \QSDW{}, respectively. \textbf{C} and \textbf{D} plot quasi-elastic RIXS intensity along $H$ and $K$ around $L=1.5$, respectively, as a function of temperature, showing the CDW Bragg peak. Error bars in \textbf{C} and \textbf{D} represent the error from Poisson statistics.}
\label{Fig1}
\end{figure*}

\dropcap{W}hen holes are doped into the Mott insulating parent compounds of the cuprates, multiple competing interactions conspire to form a rich phase diagram. In the underdoped regime, holes can save energy by clustering together on neighboring sites in order to minimize the number of broken magnetic bonds, but by doing so they pay an extra energy cost of the increased inter-site Coulomb repulsion and reduced kinetic energy. Several early theoretical works suggested that frustration between these different ordering tendencies generates an instability towards spin density wave (SDW) order \cite{Zaanen1989,Machida1989, Poilblanc1989, Emery1990, Kato1990} and low-energy incommensurate SDW correlations were indeed observed around the same time \cite{Birgeneau1989, Cheong1991, Thurston1992}. Such considerations were key to the discovery of ``stripes'' in the La$_{2-x-y}$(Nd/Eu)$_{y}$(Sr/Ba)$_{x}$CuO$_4$ or 214 family of cuprates. These correlations were found to be strongest at a doping level of $1/8$ for which static spin and charge order forms at wavevectors related by a factor of two \cite{Tranquada1995, Fujita2004}. This phase was often conceptualized in terms of a dominant spin degree of freedom, as the underdoped cuprates have a large magnetic energy scale and a relatively small electronic density of states at the Fermi level \cite{Zaanen1989,Machida1989, Poilblanc1989, Emery1990, Kato1990}. Furthermore, although high-temperature spin correlations were easily seen \cite{Cheong1991, Thurston1992, Fujita2004}, directly detecting high-temperature charge correlations proved beyond the sensitivity of standard x-ray and neutron scattering measurements. Most compellingly, charge and spin ordering appeared, until recently, to be absent in cuprates in which there was a low-energy spin gap such as YBa$_2$Cu$_3$O$_{6+x}$ (YBCO), Bi$_{1.5}$Pb$_{0.5}$Sr$_{1.54}$CaCu$_{2}$O$_{8+\delta}$ (BSCCO2212), and HgBa$_{2}$CuO$_{4+\delta}$ (HBCO1201), so the discovery of CDW correlations in these systems generated great interest \cite{Ghiringhelli2012, Chang2012, Achkar2012,Sebastian2012, Comin2014,Neto2014, Thampy2014, Hashimoto2014, tabis2014}. While the similarity of CDW phase diagrams in these materials may indicate a unified CDW mechanism \cite{Comin2016, Fradkin2015}, many of the CDW properties reported in these materials were, however, notably different than that in \LBCO{}. The CDW incommensurability in YBCO is 0.3 rather than 1/4 at 1/8 doping \cite{Ghiringhelli2012, Chang2012}, the CDW ordering seems to compete with SDW ordering  \cite{Blanco-Canosa2013,Blanco-Canosa2014, Hucker2014} and the CDW incommensurability decreases weakly with doping, rather than increasing \cite{Blanco-Canosa2014, Hucker2014,Neto2014, Comin2016}. On this basis, concepts such as nesting and electron-phonon coupling for CDW formation in YBCO, BSCCO, and HBCO were discussed extensively impeding efforts to understand these cuprates using similar mechanism that were discussed for 214 systems \cite{Comin2014,LeTacon2014,Wang2014,Comin2016,Liu2016}. Here we use new RIXS instrumentation to discover CDW correlations in the high-temperature phase of the canonical stripe-ordered cuprate \LBCO{} \cite{Fujita2004, Hucker2011, Wilkins2011, Chen2016, Achkar2016}. We find that these high-temperature CDW correlations exist without related SDW correlations at half their wavevector and that the correlations evolve with temperature, away from an incommensurability of $1/4$. These observations show that stripe order is stabilized by locking of the charge and spin correlations that occurs at low temperatures and suggest that both phenomena should be understood within the same framework. 

\section*{Results}
In this work we use Cu $L_3$ edge RIXS to achieve very high sensitivity to weak charge correlations. This works by choosing a photon energy that resonances with a Cu $2p \rightarrow 3d$ core level transition in order to enhance scattering from valence electrons, while using a spectrometer to reject the strong x-ray florescence that limits the sensitivity of traditional resonant soft x-ray scattering experiments. Figure~\ref{Fig1}A shows scattering geometry employed here. Figure~\ref{Fig1}B depicts the locations of the charge and spin ordering Bragg peaks for 214 type cuprates within the two-dimensional Brillouin zone labeled $\QCDW{}\approx(0.24, 0)$ and $\QSDW{}\approx(0.38, 0.5)$ in reciprocal lattice units (r.l.u.). We start by choosing $\sigma$ polarized incident x-rays in order to enhance our sensitivity to charge scattering \cite{Ament2011, Ghiringhelli2012, Dean2015}. Figure \ref{Fig1}C,D plots projections of the quasi-elastic scattering intensity around \QCDW{} along the $H$ and the $K$ directions. A clear peak is observed at the base temperature corresponding to the known CDW with a wavevector of $(0.235, 0)$ and a correlation length of 207(5)~\AA{} \cite{ Wilkins2011, Hucker2011, Achkar2016}. As established in several previous studies the low-temperature CDW peak intensity drops with increasing temperature and seems to disappear around 55~K \cite{Wilkins2011, Hucker2011, Achkar2016}. 

%
\begin{figure*}[tb]
\centering
\includegraphics[width=16 cm]{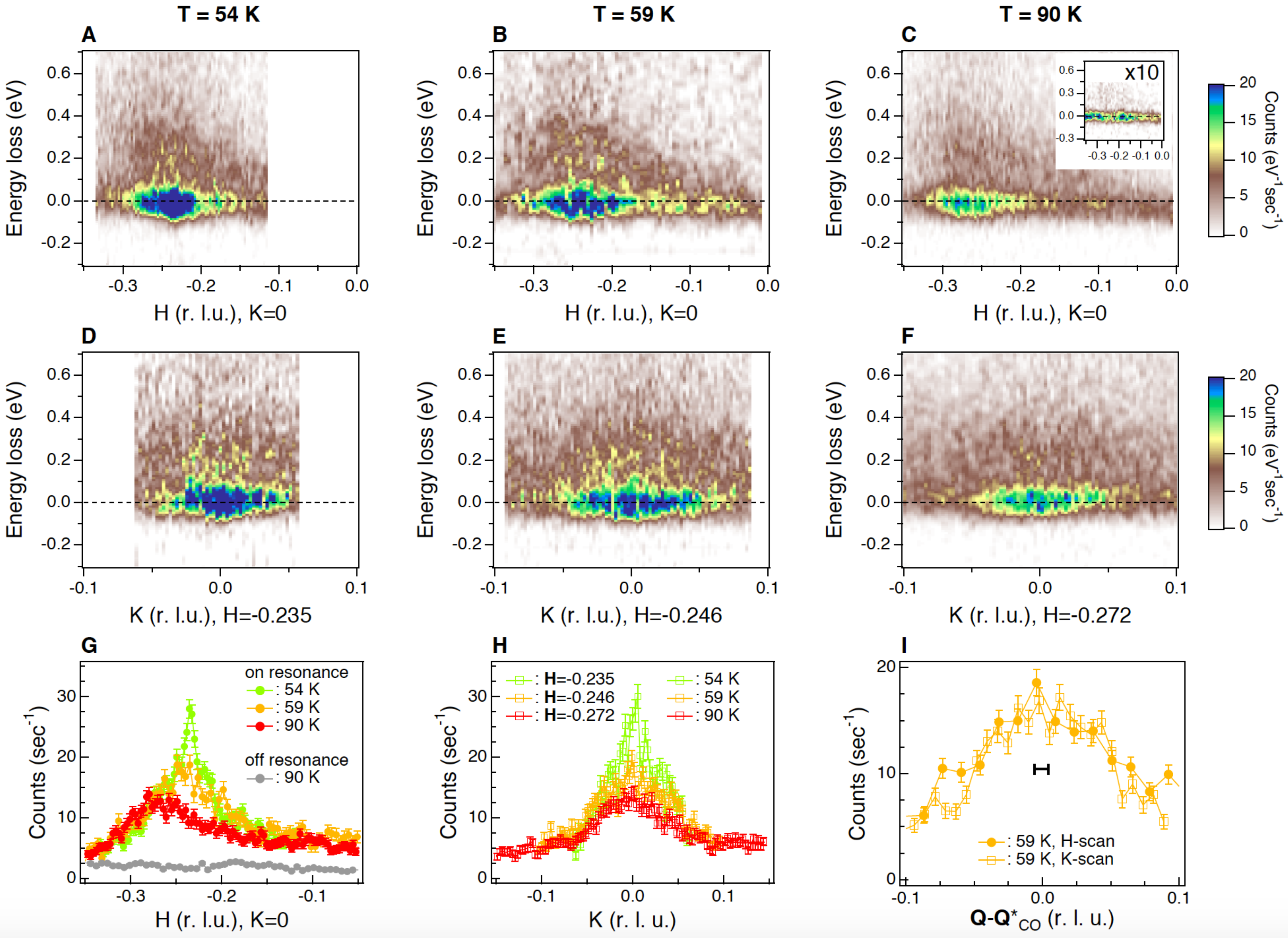}
\caption{\textbf{Identification of the high-temperature-CDW}. \textbf{A}-\textbf{F} RIXS intensity at 54, 59 and 90~K cutting through the observed peak in the quasi-elastic intensity as a function of $H$ (\textbf{A}-\textbf{C}) and $K$ (\textbf{D}-\textbf{F}). A peak in the quasi-elastic intensity is seen in the vicinity of \QCDW{} alongside an increase in the inelastic intensity. The inset of panel \textbf{C} displays an intensity map at 90~K taken with a different off-resonant x-ray energy in order to reduce the sensitivity to the valence electrons. This was multiplied by a factor of 10 to make the signal visible on the same color scale. \textbf{G},\textbf{H} The quasi-elastic intensity calculated by integrating (\textbf{A}-\textbf{F}) confirming the presence of the peak. \textbf{I} Comparison of scans in the $H$ and $K$ directions showing similar widths parallel and transverse to the CDW, similar to the low-temperature behavior (Fig.~\ref{Fig1} \textbf{C},\textbf{D}). As discussed in \textit{Results}, this scattering demonstrates the presence of high-temperature CDW correlations. Error bars in \textbf{G}-\textbf{I} come from Poisson counting statistics.}
\label{Fig2}
\end{figure*}

%
\begin{figure*}[tb]
\centering
\includegraphics[width=17 cm]{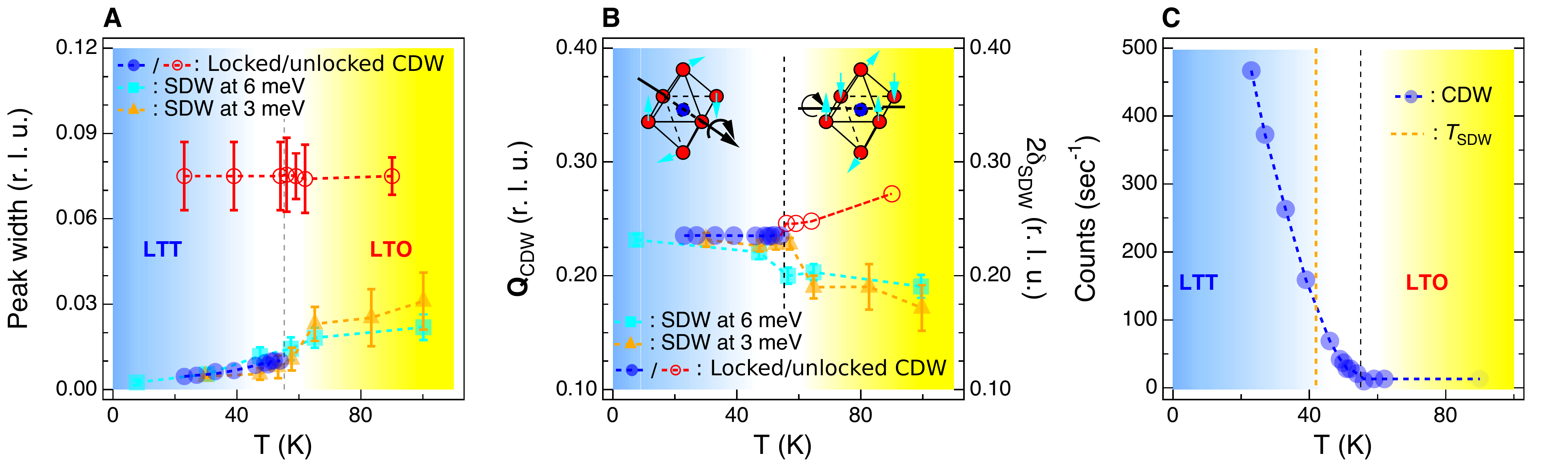}
\caption{\textbf{Decoupling of the CDW and SDW in the high-temperature phase.} \textbf{A}-\textbf{C} The results of fitting the quasi-elastic intensity showing: \textbf{A} the full width at half maximum, \textbf{B} the incommensurability and \textbf{C} the intensity at the peak. The black dashed line at 54~K corresponds to the low-temperature tetragonal (LTT) to low-temperature orthorhombic (LTO) structural phase transition which is depicted in \textbf{B}, blue and yellow code temperatures below and above this threshold \cite{Wilkins2011}. The orange dashed line at 42~K in \textbf{C} represents the static SDW transition. The behavior of the SDW, taken from inelastic neutron scattering results at 3 and 6~meV energy transfer from Ref.~\cite{Fujita2004} are included on panels \textbf{A} and \textbf{B}. We see that the CDW and SDW incommensurabilities evolve in different directions above 54~K, which indicates a decoupling of the charge and spin degrees of freedom. We also note that the high-temperature CDW width and intensity show no detectable changes through the LTT-LTO transition (any possible changes would be smaller than our error bars, which are obtained from the least-squares fitting algorithm).}
\label{Fig3}
\end{figure*}

Having established the low-temperature CDW properties, we scanned large regions of reciprocal space at temperatures above the nominal transition. Figure~\ref{Fig2}A-F plots RIXS intensity maps that reveal broad momentum-dependent scattering. The quasi-elastic intensity in these maps show broad peaks around $(H, K)=(0.24,0)$ for temperatures of 54-59~K, while at higher temperatures it peaks at larger $H$ while remaining centered at $K=0$. Although the close match in the wavevectors between the low-temperature and high-temperature scattering already indicates an intimate connection between this scattering and the low-temperature CDW, it is important to justify the electronic origin of the broad peak. Panels~\ref{Fig2}C (inset) and \ref{Fig2}G show the off-resonance RIXS intensity map and integrated RIXS intensity along $H$. Both the inelastic excitations and the quasi-elastic remnant intensity are significantly suppressed when changing the incident energy, proving that the signal is dominated by the x-ray resonant process. The flat 2.8(0.3) counts sec$^{-1}$ off-resonant intensity also confirms the constant spectrometer acceptance.  We also see in Fig.~\ref{Fig2}I that the peak has the same width in $H$ and $K$, consistent with the behavior of the low-temperature CDW (Fig.~\ref{Fig1}C,D). It is also worth noting that x-ray self-absorption effects (see Fig.~S2) and the Cu $L_3$ RIXS cross section are known to vary monotonically in this scattering geometry \cite{Ament2011,  Ghiringhelli2012}. Based on all these experimental observations, we conclude that the observed broad peak represents a direct observation of the high-temperature CDW correlations discussed extensively ever since the discovery of the low-temperature CDW \cite{Tranquada1995, Kivelson2003, Vojta2009,Nie2014,Capati2015}.\footnote{We chose the term ``high-temperature CDW correlations'' as the most generic way to refer to valence charge modulations with a different periodicity to the underlying lattice.}\ Compared to the low-temperature CDW, the high-temperature CDW has far lower peak intensity (13 vs 467 counts/s) but a much broader line width (about a factor of 16). As a result, this diffuse high-temperature scattering comprises approximately 7 times larger 2D-momentum and energy integrated spectral weight than the sharp low-temperature CDW peak that emerges on top of the diffuse scattering below 54~K. 

Upon cooling through the 54~K transition, no changes are observed in the diffuse tail of intensity. Although this is opposite to what is expected in a disorder-free phase transition, in which all high-temperature correlations would be expected to condense into a sharp CDW peak, such behavior is expected in the presence of disorder \cite{Chatterjee2015, Nie2017}. Cuprates are known to host appreciable disorder \cite{Alloul2009, Hucker2010, Hucker2014, Campi2015}, and this is the likely cause of the observed phenomenology, particularly in view of the match between the structural and CDW correlation length in \LBCO{} under pressure \cite{Hucker2010}.

High-temperature CDW correlations in 214 cuprates are often argued to be dynamic \cite{Kivelson2003, Reznik2006, Vojta2009} and such a view is supported by transport measurements \cite{Li2007}. Notably, long-range ordered static LTT octahedral tilts, often thought to be coupled to the CDW at low temperatures, are found to become dynamic and correlated over a $\sim 10$~\AA{} length scale above the transition similar to the high-temperature CDW correlation length detected here \cite{Bozin2015,Fabbris2013}. On the basis of the resolution-limited energy width we observe (see Fig.~S4), we conclude that the high-temperature CDW is static on a timescale of $\sim$100~fs, but slow fluctuations are nonetheless possible and can, in principle be directly measured by other techniques \cite{Chen2016}. 

Figure~\ref{Fig3} shows the temperature dependence of the CDW correlations as determined by fitting Lorentzian-squared functions to the diffuse CDW peak intensity present at all temperature and the sharp CDW peak that emerges at low temperatures (see Fig.~S5 and Fig.~S6). The correlation length of the high-temperature CDW of 13(2)~\AA{} is much shorter than that in the low-temperature state  (207~\AA{}) and substantially shorter than  YBCO (60~\AA) \cite{Ghiringhelli2012}, but is of the same order of magnitude as several other cuprates systems such as  Bi$_2$Sr$_{2-x}$La$_x$CuO$_{6+\delta}$ (BSLCO2201) (12~\AA{}) \cite{Comin2014}, BSCCO2212 ($<$24~\AA{}) \cite{Hashimoto2014}, La$_{2-x}$Sr$_x$CuO$_4$ (LSCO) (35~\AA{}) \cite{Thampy2014}, HBCO1201 (20~\AA{}) \cite{tabis2014}, hinting that the high-temperature CDW properties may help reconcile the difference between different cuprates. Further clues are evident in the wavevecector behavior allowing to associate the high temperature state observed here to the low temperature behaviour of other compounds. As can be seen in Fig.~\ref{Fig3}B, for temperatures below 55~K the incommensurability of the CDW and SDW appear to be locked by a factor of two, which is a well-known property of 214-type cuparates \cite{Cheong1991, Thurston1992, Tranquada1995, Hucker2011, Fujita2004}. Upon heating above 55~K, we see strong violation of this relation [Fig.~\ref{Fig3}B]: The CDW correlations evolve away from $H\approx 1/4$ and away from twice the incommensurability of the SDW (i.e.\ the CDW and SDW decouple) \cite{Fujita2004}. 

\begin{figure*}[tb]
\centering
\includegraphics[width=12 cm]{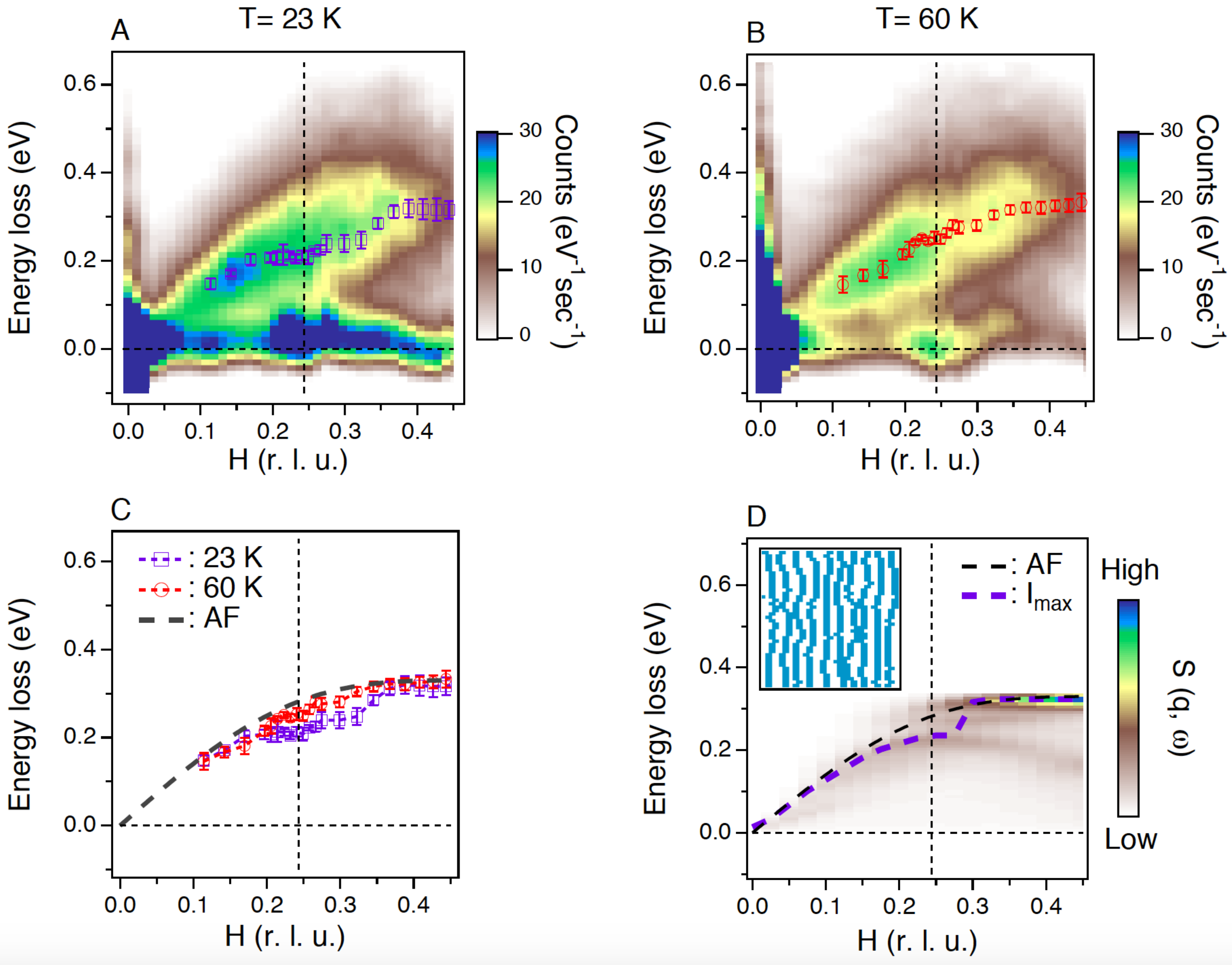}
\caption{\textbf{Magnetic excitation spectrum and charge-spin coupling.} \textbf{A},\textbf{B} RIXS intensity maps measured in a geometry that couples to the paramagnon excitation at 23~K in the low-temperature CDW phase and at 60~K in the high-temperature CDW phase. Purple squares and red circles are the extracted peak positions from fitting the paramagnon lineshape (see Fig.~S7). \textbf{C} Comparison of the peak positions obtained by fitting the data in \textbf{A} and \textbf{B} to the dispersion expected from spin wave theory (SWT) in an antiferromagnet (AF) without stripes. \textbf{D} The inset shows the theoretical charge configuration for one realization of the meandering stripes which mimics the experimental charge structure factor at 23~K. Stripes with a width of two sites with increased hole concentration are shown in blue. The main panel shows an average of the magnetic dynamic structure factor for ten such configurations as described in the method section. The purple dashed curve connects the points of maximum intensity at each $H$ value. Such a picture qualitatively captures the observed dispersion in \textbf{C}.}
\label{Fig4}
\end{figure*}

We further tested the nature of the CDW/SDW state and its charge-spin coupling by changing the RIXS geometry in order to measure the magnetic excitation spectrum in the same $\mathbi{Q}$-range (see Fig.~S1) \cite{Ament2011}. Inelastic neutron scattering has been applied extensively to study the magnetic excitations around $\mathbi{Q}_\text{SDW}$, finding an `hour-glass' shaped dispersion \cite{Tranquada2004, Seibold2006, Vojta2009}, RIXS can study the magnetic spectrum around $\mathbi{Q}_\text{CDW}$, a region of reciprocal space in which stripe-related effects have never been observed. Figure~\ref{Fig4}A-B shows the resulting spectral intensity above that is dominated by damped spin wave excitations called paramagnons \cite{LeTacon2011,Dean2012, DeanLSCO2013,DeanBSCCO2013}. We fit the paramagnon dispersion (see the SI for more details) and are compared it to the dispersion expected for a standard N\'{e}el antiferromagnet (AF) in Fig.~\ref{Fig4}C, finding a softening of the excitation energy over a broad range of reciprocal space around \QCDW{}. The significant deviation observed at low temperatures shows that stripe-formation modifies the short-range spin correlations around \QCDW{} at low temperature, but this coupling is much reduced at higher temperatures, consistent with a weakened charge-spin coupling above the transition. There have been extensive efforts to model such stripe-related modifications in the spin excitation spectrum as this provides a means to develop detailed models for the character of the ground state \cite{Carlson2004,Seibold2005,Seibold2006,YaoPRB2006,Vojta2009,Lorenzana2002,Seibold2012}. These theories do a good job of capturing the magnetic dispersion around \QSDW{}, but none of these theories adequately capture the dispersion around \QCDW{}. We discovered that a partially ordered CDW state with meandering charge stripes (see the inset of Fig.~\ref{Fig4}D), as constrained by the measured charge scattering, does successfully capture the observed modification in the magnetic dispersion. Figure~\ref{Fig4}D plots our calculations (see the methods section for full details). Despite the simplicity of the model, it captures what is observed in Figure~\ref{Fig4}C, confirming that we have identified the essential features of the ground state. Calculations based on a perfectly stripe-ordered crystalline CDW predict several sharp modes which are not observed (see Fig.~S8). 

%

\section*{Discussion}
Our results have important implications for the relationship between stripe order in 214 type cuprates and CDW order in non 214 cuprates \cite{Keimer2015}. We show that the high temperature state of \LBCO{} hosts CDW correlations at a wavevector unlocked from the SDW wavevector. This establishes an appealing analogy to non-214 systems, which also host CDW correlations without any obviously related SDW correlations. Indeed, stripe order in \LBCO{} appears to form via locking of the CDW and SDW at low temperatures. A remaining discrepancy, however, is that pristine non-214 systems tend to exhibit a spin gap not present in 214 systems \cite{Hinkov2007, Stock2005, Stock2010, Xu2009}. Substituting 2\% Zn for Cu in YBa$_2$Cu$_3$O$_{6.6}$ is known to close the spin gap and stabilize SDW order with an incommensurability of 0.1, but this remains unlocked from the CDW incommensurability of 0.3 \cite{Suchaneck2010}.  We furthermore demonstrate that the wavevector of the high-temperature CDW correlations in \LBCO{} is not uniquely defined by the doping level contrary to what is seen in the low temperature state \cite{Cheong1991, Thurston1992, Tranquada1995, Hucker2011, Fujita2004}. Indeed the high-temperature wavevector of $\mathbi{Q}_\text{CDW}= 0.272(2)$ is closer to what is seen in $1/8$ doped YBCO and BSCCO2201, which are 0.32 and 0.27, respectively than in the low temperature wavevector of 0.235 \cite{Comin2014, Ghiringhelli2012}. Temperature dependent wavevectors have been predicted in Landau-Ginzberg modeling of stripe ordering scenarios \cite{Zachar2000, Nie2017}. In these models, the CDW wavevector is determined by competition between the CDW's intrinsic ordering wavevector and coupling between the CDW and another degree of freedom, such as the SDW. The low-temperature wavevector may consequently not reflect the formation mechanism for these phases. Another unresolved discrepancy is that that low-temperature CDW incommensurability in \LBCO{} increases with temperature \cite{Fujita2004, Hucker2011}, distinct from the weak decrease in CDW incommensurability seen in YBCO and other non-214 systems \cite{Blanco-Canosa2013, Blanco-Canosa2014, Hucker2014}. It will be important for future studies to measure whether the high-temperature CDW correlations in La$_{2-x}$Ba$_{x}$CuO$_{4}$ increase or decrease with $x$. Here we show that the CDW incommensurability in \LBCO{} changes by 0.04~r.l.u.\ (from 0.235 to 0.272) with temperature i.e.\ on thermal energy scale of order 10~meV. We note that this is about the same magnitude as the doping induced change in YBCO (from 0.34 at 0.09 hole concentration to 0.30 at 0.16) \cite{Blanco-Canosa2014, Hucker2014}.

We end by discussing different ways to reconcile the phenomenology observed here with that seen in other cuprate materials. One option is to assume that 214 and non-214 cuprates host completely different types of CDW. Alternatively, one can posit a universal CDW formation mechanism in which many states with different ordering wavevectors and different inter-plane stacking configurations can exist with only small energy differences. The widely discussed strongly correlated mechanisms are examples of this as the wavevector is determined by a balance of different competing interactions \cite{Zaanen1989, Vojta2009, Machida1989, Poilblanc1989, Kato1990, Emery1990, Kivelson2003, Castellani1995, Lorenzana2002, Dodaro2017}, in contrast to nesting in which the wavevector is expected to correspond to parallel features in the Fermi surface \cite{Johannes2008}. In this universal scenario, low-temperature ordering wavevector in 214 systems would then be defined by coupling between the CDW and the SDW condensing a relatively small fraction of the available low-energy fluctuations together into well correlated CDW order and establishing the factor of two relationship between the CDW and SDW incommensurabilities. In non-214 systems this mechanism does not occur due to the spin gap and the absence of the low-temperature tetragonal (LTT) structure, which is believed to play an important role to stabilize the CDW \cite{Tranquada1995}. In this case other details may be relevant for determining the low-temperature wavevector. Several researchers have pointed to analogies between Fermi surface features and the CDW wavevector in this case \cite{Ghiringhelli2012, Comin2014,Neto2014, tabis2014,Neto2016}. A common origin for CDWs correlations in all cuprates also naturally explains why the onset temperature peaks at around $1/8$ doping in all families. 

\section*{Conclusions}
We exploited the high sensitivity of RIXS to discover charge correlations in the high temperature state of \LBCO{}. These correlations show that La-based 214 type cuprates can host CDW correlations that are unlocked from the SDW suggesting stripes form via the locking of the charge and spin wavevectors at low temperatures. This establishes shared properties between different cuprates, constraining models for the normal states from which high-temperature superconductivity emerges. 

\matmethods{A \LBCO{} single crystal was grown using the floating zone method and cleaved ex-situ to reveal a face with a $[001]$ surface normal. The wavevectors used here are described using the high temperature tetragonal ($I4/mmm$) space group with $a = b = 3.78$~\AA{} and $c = 13.28$~\AA{}. Correlation length is defined as $1/\text{HWHM}$ where $\text{HWHM}$ is the half width at half maximum of the peak in reciprocal lattice units. 

\noindent RIXS measurements were performed at the ID32 beamline of the European Synchrotron Radiation Facility (ESRF). The resonant condition was achieved by tuning the incident x-ray energy to the maximum of the Cu $L_{3}$ absorption peak around 931.5~eV. The scattering geometry is shown in Fig.~\ref{Fig1}A. $\sigma$ and $\pi$ x-ray polarizations are defined as perpendicular and parallel to the scattering plane, respectively. $H$ and $K$ scans are achieved by rotating the sample around the $\theta$ and $\chi$ axes, without changing $2\theta$, thus changing the in-plane component of the momentum transfer $\mathbi{Q}$ = $\mathbi{k}_{f}-\mathbi{k}_{i}$. By doing this, we are assuming that the scattering is independent of $L$, which is reasonable as the inter-layer coupling in the cuprates is known to be weak \cite{Wilkins2011, Hucker2011, DeanLSCO2013}. Positive (negative) $H$ corresponding to larger (smaller) $\theta$ values. The horizontal and vertical momentum resolution was 0.008~\AA$^{-1}$ and 0.001~\AA$^{-1}$, respectively and all intensities are normalized to beam current and counting time. Two different geometries are used here to provide sensitivity to charge and spin degrees of freedom respectively \cite{Ament2011, Ghiringhelli2012}. For the charge scattering, we used $\sigma$-polarized incident x-rays and negative $H$ values. The spectrometer scattering angle (2$\theta$) was fixed at $118^{\circ}$ such that $L\approx 1.5$ and the total instrumental energy resolution (full-width at half maximum) was set to 90~meV to increase the counting rate.  The quasi-elastic intensity was obtained by integrating the RIXS spectrum in an energy window of $\pm$ 150~meV around 0~meV. To measure the spin excitation spectra, we used $\pi$-polarized incident x-rays and positive $H$ values. The scattering angle (2$\theta$) was set at the maximum value of $149^{\circ}$ to access higher $H$ values and the total instrumental energy resolution was set to 70~meV. The elastic energy was determined by measuring the diffuse scattering from carbon tape for every spectrum obtained. 

\noindent We performed our calculations of the spin excitation spectrum in the CDW state starting with an initially ordered set of charge stripes on a $40 \times 40$ site lattice. We then used a Monte Carlo algorithm to disorder the stripes until the charge structure factor matches the measured CDW peak shape and computed the magnetic excitation spectrum of the disordered state using a suitably parametrized Heisenberg model \cite{Carlson2004}. It is assumed that the charge stripes define domain walls across which magnetic exchange $J$ is replaced by a ferromagnetic exchange $J_F$. $J=165$~meV was chosen to match our observed zone boundary magnon energy and $J_{F}=-0.09J$ was chosen to obtain the correct energy for the neck of the hourglass in Ref.~\cite{Tranquada2008}. The magnetic excitation spectrum of the system with the domain walls is computed under the spin wave theory (SWT) approximation and averaged over the expected different domain configurations. Such a treatment is sufficient to reproduce the observed magnetic peak dispersion even without including other effects such as fermionic excitations \cite{Seibold2006, Vojta2009}.

}

\showmatmethods{}

\acknow{We thank E. Bozin, C. Mazzoli, T. M. Rice, J. Tranquada and S. Wilkins for discussions and J. Pelliciari for assistance. H.M.\ and M.P.M.D.\ are supported by the Center for Emergent Superconductivity, an Energy Frontier Research Center funded by the US Department of Energy (DOE), Office of Basic Energy Sciences. Work at Brookhaven was supported by the U.S.\ DOE (Contract No.~DE-SC00112704). Theoretical work by J.L.\ is supported by the Italian MIUR (project PRIN-RIDEIRON-2012X3YFZ2). The experiment was performed at ID32 at the ESRF.}

\showacknow{} 
\bibliography{refs}

\end{document}